\documentclass[10pt,reqno,oneside]{amsart}
\usepackage{verbatim,amsmath,amssymb,cite,epsfig,subfigure,color,hyperref,ulem}

\numberwithin{equation}{section}

\newcommand{\tr}{\operatorname{tr}}

\newcommand{\sech}{\operatorname{sech}}

\newcommand\smallO{
  \mathchoice
    {{\scriptstyle\mathcal{O}}}
    {{\scriptstyle\mathcal{O}}}
    {{\scriptscriptstyle\mathcal{O}}}
    {\scalebox{.7}{$\scriptscriptstyle\mathcal{O}$}}
  }

\begin{document}

\title[Non-Equilibrium Thermodynamics And Statistical Mechanics in Tornado Theory]
{Review of Non-Equilibrium Thermodynamics And Statistical Mechanics of Vortex Gases in Tornado Theory}

\author{Pavel B\v{e}l\'{\i}k}
\author{Douglas Dokken}
\author{Mikhail Shvartsman}

\address{P.~B\v{e}l\'{\i}k\\
Department of Mathematics, Computer Science, and Data Science\\
Augsburg University\\
2211 Riverside Avenue\\
Minneapolis, MN 55454\\
U.S.A.}
\email{belik@augsburg.edu}

\address{D.~Dokken\\
Department of Mathematics\\
University of St.~Thomas\\
2115 Summit Ave.\\
St.~Paul, MN 55105\\
U.S.A.}
\email{dpdokken@stthomas.edu}

\address{M.~Shvartsman\\
Department of Mathematics\\
University of St.~Thomas\\
2115 Summit Ave.\\
St.~Paul, MN 55105\\
U.S.A.}
\email{mmshvartsman@stthomas.edu}


\keywords{
Quasi-Two-Dimensional Turbulence;
Thermodynamic Fluxes;
Entropy;
Non-Equilibrium Thermodynamics;
Nonlinear Schr\"{o}dinger Equation;
Gross--Pitaevskii Equation;
Madelung and Hasimoto Transforms;
Vortex Stretching.}

\subjclass[2010]{35Q30, 35Q31, 35Q55, 76D05, 76F40, 80A17, 94A17}

\date{\today}

\begin{abstract}
 This work puts into mathematical, statistical mechanical, and thermodynamical context the initial stages of the genesis of tornado-like vortices with the aim to be consistent with the current state of knowledge of the process of tornadogenesis. In particular, it discusses a mathematical foundation of the formation of coherent structures such as ``cusps'' and ``hairpins'' using variants of the nonlinear Schr\"odinger equation that arise via the Hasimoto transform of a vortex filament model. The behavior of such structures is then analyzed within a quasi-two-dimensional boundary layer model using the statistical mechanics of vortex gases to explain the rearrangement of cusps and other vertical vortex filaments into patches and possibly supercritical vortices. Non-equilibrium thermodynamics is used to obtain the entropic balance and the internal entropy production rate, and connect them to the turbulent heat flux. A formula for the non-equilibrium turbulent heat supply and formulas for the entropy supply and entropy production in the boundary layer are also provided. A relationship between the vorticity and the entropy gradient based on macroscopic fluctuations is given with implications to stretching and tilting of vorticity in the vertical direction. We conclude with some remarks on equivalence of Schr\"odinger and Gross--Pitaevskii equations in describing vortex filaments.
\end{abstract}

\allowdisplaybreaks
\thispagestyle{empty}
\maketitle

\section{Introduction}
\label{sec:introduction}
A recent survey paper by Fischer et al.~\cite{fischer24} has outlined the current state of knowledge about the process of tornadogenesis, reducing the process to a conceptual model with four stages. They are:
\begin{enumerate}
    \item Mesocyclone formation, with the associated pressure perturbations driving upward acceleration and stretching near and below cloud base (critical for Stage 3 below).
    \item Generation of vortex patches (amorphous surface-level vertical vorticity not yet with vortex characteristics), in simulations predominantly via reorientation of baroclinically generated vorticity in downdrafts.
    \item Organization of one or multiple vortex patches to a single symmetric vortex and intensification via stretching.
    \item Transition to a developed tornado with a vortex boundary layer and corner flow region in which horizontal vorticity is tilted into the vertical right above the surface.
\end{enumerate}
Observational and numerical motivation for this conceptual model are discussed in the survey paper and additional problems that need further study or are unresolved are listed therein. Specifically, the organization and intensification of the pre-tornadic vorticity near the surface and its rapid development upward during Stage 3 is emphasized, together with the uncertainty of the 3D evolution of these vortex patches. The question of the role and importance of the streamwise vorticity current and surface roughness in Stage 1 is raised, together with the role of turbulent eddies in the environmental boundary layer \cite{markowski24}. Other questions relate to the processes or mechanisms that determine the tornado size and intensity, both on the supercell scale (Stages 1 and 2) and on the vortex scale (Stages 3 and 4), as well as to the importance of supercell-external factors, such as storm mergers and mesoscale boundaries.

A recent computational paper by Parker \cite{parker23} argues that two phenomena are important for the generation of tornado-like vortices. A careful study is performed and it is concluded that a sufficient amount of unstructured surface vertical vorticity, $\zeta$, when coupled with a strong enough updraft, are sufficient for the formation of a tornado-like vortex. In particular, quasi-random $\zeta$-fields are generated via the use of Perlin noise \cite{perlin85}, and an updraft nudging technique \cite{naylor12} is used. It is concluded that regardless of the origin of the vertical vorticity, tornado-like vortices get produced when enough initial surface circulation and a large enough vertical velocity vertical gradient ($\partial w/\partial z$ in the standard notation) near the ground are present. The former aspect would correspond to Stage 2 and the latter aspect to the intensification via stretching in Stage 3.

Investigation of atmospheric flows, especially the ones related to tornado-like events, requires a significant contribution from thermodynamic considerations, since a significant portion of a thunderstorm's energy is generated by latent heat release associated with water vapor condensation. Citing \cite{rotunno15}: {\it``There is a strong nexus with thermodynamics, because these thunderstorms are driven by the phase change of water vapor. There are lots and lots of things other than pure fluid dynamics in this field.''}

To bypass the complications of storm thermodynamics, researchers and forecasters utilize various indices, in particular, combined mechanical and thermodynamic parameters, such as the fixed layer significant tornado parameter (STP). The STP is a composite index that includes the 0--6 km bulk wind difference (6BWD), the 0--1 km storm-relative helicity (SRH1), the surface based parcel convective available potential energy (sbCAPE), and the surface based parcel lifting condensation level (sbLCL). It is defined in the following way:
\begin{equation*}
    \text{STP} = \frac{\text{sbCAPE}}{1500} \frac{\text{J}}{\text{kg}} \frac{2000-\text{sbLCL}}{1000\text{ m}}
    \frac{\text{SRH1}}{150} \frac{\text{m}^2}{\text{s}^2} \frac{\text{6BWD}}{20} \frac{\text{m}}{\text{s}}.
\end{equation*}
The sbLCL is set to 1.0 when sbLCL $<$ 1000 m, and it is set to 0.0 when sbLCL $>$ 2000 m; the 6BWD term is capped at a value of 1.5 for 6BWD $>$ 30 m/s, and set to 0.0 when 
6BWD $<$ 12.5 m/s \cite{stp}.

A majority of significant tornadoes (EF2 or stronger on the Enhanced Fujita scale \cite{efscale}) have been associated with STP values greater than 1, while most non-tornadic supercells have been associated with values less than 1 in a large sample of Rapid Refresh (RAP) analysis proximity soundings. The RAP is a continental-scale NOAA hourly-updated assimilation modeling system operational at the National Center for Environmental Prediction (NCEP) \cite{rap}. It covers North America and is comprised primarily of a numerical forecast model and an analysis\slash assimilation system to initialize that model. The definitions of the updated indices are given on the NOAA website \cite{indices}. A detailed discussion of the combined and other tornado indices is given in \cite{doswell06, grams12}. In particular, it is stated in \cite{doswell06}: {\it``Forecasters and researchers are seeking a `magic bullet' when they offer up yet another combined variable or index for consideration.''}

In this paper we develop ideas from non-equilibrium thermodynamics and statistical mechanics of vortex gases in the context of a turbulent boundary layer, and we investigate how these ideas are related to the various stages of tornadogenesis outlined above and to the findings in \cite{parker23}. We describe how the formation of coherent structures, such as cusps along an originally nearly straight horizontal vortex, can be modeled by the solutions of nonlinear Schr\"odinger equations. We will use the statistical mechanics of vortex gases to investigate the development of pre-tornadic vorticity that is first organizing and intensifying near the surface, and also the development of vortex patches and turbulent eddies in the environmental boundary layer. We use non-equilibrium thermodynamics to study the development of conditions in the boundary layer facilitating rapid vertical development of vorticity from the boundary layer to the cloud base. One of the aims of this work is to establish a consistent thermodynamic approach aiming to bypass the ad-hoc nature of thermodynamic indices and develop a consistent model of non-equilibrium thermodynamic evolution in a tornado-like flow. We attempt to develop a better understanding of how non-equilibrium thermodynamics accounts for low-level boundary turbulence that leads to the generation of tornado-like vortices and their subsequent stretching.

The case of three-dimensional turbulence in a shallow layer that can be approximated as two-dimensional is called quasi-two-dimensional turbulence. We use this term later in Section~\ref{sec:statmech} to describe the situation studied there. In two-dimensional flows, energy $E=\frac12\langle|{\bf v}|^2\rangle$ and enstrophy $\Omega=\frac12\langle|{\boldsymbol{\omega}}|^2\rangle$ are conserved quantities. In these formulas, $\bf v$ and $\boldsymbol\omega$ are the velocity and vorticity fields, respectively, and $\langle\ldots\rangle$ denotes their space averages. The energy spectrum, $E(k)$, and the enstrophy spectrum, $\Omega(k)$, are related via $\Omega(k) = k^2 E(k)$, where $k$ is the wave number, which implies that in two dimensions the energy must cascade to larger scales and the enstrophy must cascade to smaller scales \cite{alexakis23, batchelor69, kraichnan67, kraichnan71, kraichnan75}. The extent to which this fails to hold in three dimensions is related to whether turbulence in thin surface layers is quasi-two-dimensional or not \cite{alexakis23, kraichnan75}.

\begin{figure}
    \begin{center}
        \includegraphics[width=0.4\textwidth]{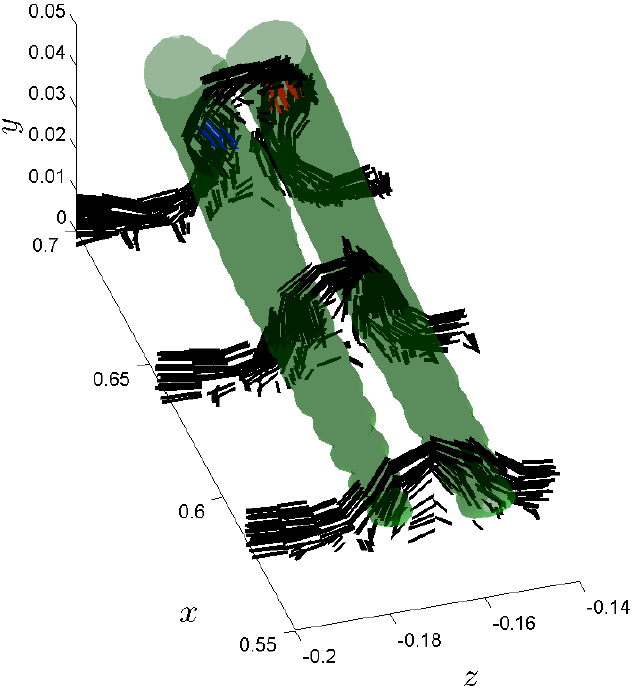}
        \hfill
        \includegraphics[width=0.55\textwidth]{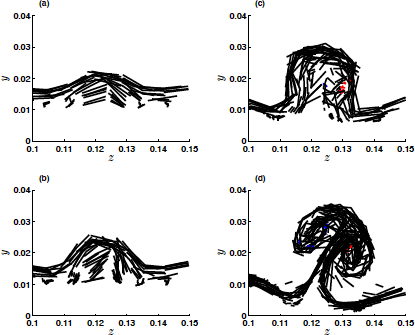}
    \end{center}
    \caption{Development of vortex filaments in the flow past a thin horizontal plate~\cite{bernard11}. The onset of furrows and cusps, along with the green isosurfaces of rotation (left); filaments in the furrow viewed by a moving observer (right). Filaments closely aligned with the $x$-direction are colored red and blue.}
    \label{fig:bernard1}
\end{figure}

\begin{figure}
    \begin{center}
        \includegraphics[width=0.5\textwidth]{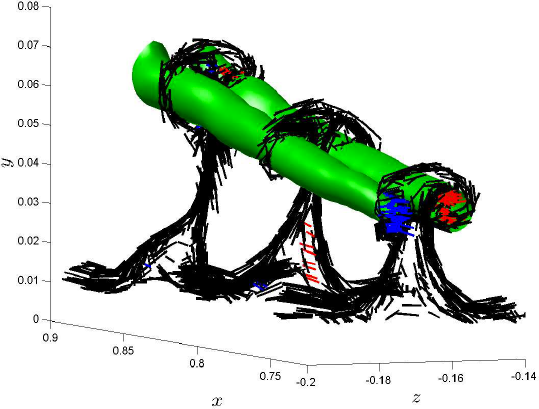}
        \hfill
        \includegraphics[width=0.45\textwidth]{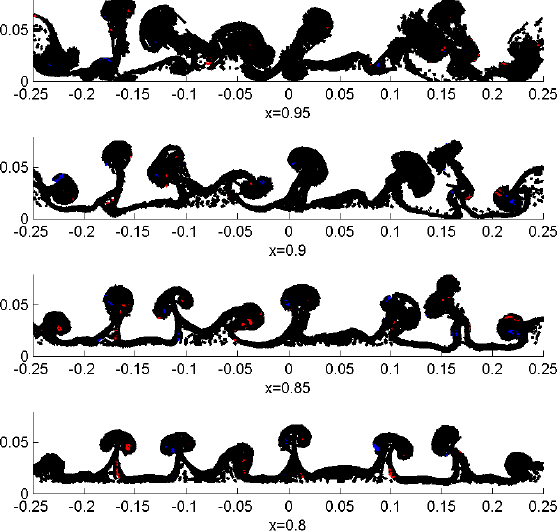}
    \end{center}
    \caption{Isosurfaces of rotation (green) and the mushroom-like cusps (left); filaments at a fixed time after mushroom-like cusps have developed and before the onset of a turbulent flow (right).}
    \label{fig:bernard2}
\end{figure}

\begin{figure}
    \begin{center}
        \includegraphics[width=0.85\textwidth]{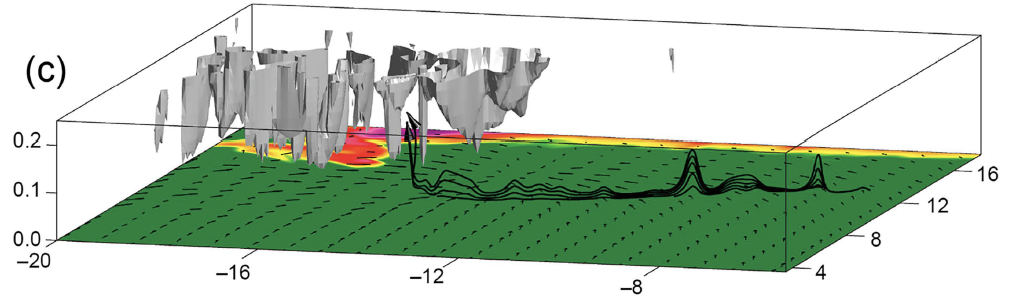}
    \end{center}
    \caption{Isosurfaces of the vertical velocity $w=3$ m s$^{-1}$ (grey) and near surface vortex lines passing through $(x,y)=(-9, 12)$ at $z=7.5$, $12.5$, $17.5$, $22.5$, and $27.5$ m. A continuum of these vortex lines would form a corrugation of vertical vorticity as seen in \cite{markowski24} and furrows and cusps consistent with the results in \cite{bernard11}.}
    \label{fig:coherent}
\end{figure}

Using a hybrid vortex filament scheme, Bernard \cite{bernard11} studies a boundary layer within a flow past a thin horizontal plate and the subsequent development of turbulence. The near-surface horizontal vortex filaments originally transverse to the flow develop perturbations referred to as furrows. This is shown in the left image of Fig.~\ref{fig:bernard1} (see also Figs.~1 and 4 in \cite{bernard11}). The initially forward-tilting arch-like filaments over time turn into mushroom-like arrangements as seen in the right images of Fig.~\ref{fig:bernard1} and in Fig.~\ref{fig:bernard2}. The rotational part of the vorticity is captured by visualizing the $\lambda_2$-isosurfaces of rotation, shown as green ``hairpin vortex legs'' in Figs.~\ref{fig:bernard1} and \ref{fig:bernard2}. The filament segments create cusps between the hairpin legs, and the hairpin legs eventually pass through the lobes of the mushroom caps as seen in the left image of Fig.~\ref{fig:bernard2}. The vorticity within the hairpin legs is mostly aligned with the streamwise direction of the legs and is shown by the red and blue color coding in Figs.~\ref{fig:bernard1} and \ref{fig:bernard2}. The hairpin legs can occur as counter-rotating pairs as shown, or singly (not shown in this paper), and it is emphasized in \cite{bernard11} that {\it``there is much non-rotational vorticity engaged in the functioning of the vortical structures that produce hairpins''} and that neglecting the role of non-rotational vorticity has led to {\it``the illusion that hairpin vortices and packets are structures in their own right''}. Consequently, the formation of the cusps appears to be an essential mechanism leading to turbulence and possible eventual tornadogenesis, which is supported by computational evidence in \cite{markowski24} and shown in Fig.~\ref{fig:coherent}. For a shallow surface layer, the behavior of the cusps can be approximated by a quasi-two-dimensional model. Simple examples of how surface friction can produce separation in the boundary layer, and thus vorticity, are given in \cite{chorinmarsden92} (section 2.2); see also \cite{adrian07,bernard11,bernard19}.

Given the quasi-two-dimensional nature of the cusp and hairpin vortices, they do not dissipate \cite{alexakis23}. Instead, an inverse energy cascade forms to produce coherent (turbulent) structures that are later subject to vortex stretching and may produce a tornado \cite{orf17}. Due to the inverse energy cascade, the filaments concentrate into clusters of vorticity that satisfy the cyclostrophic balance conditions. The breakdown of the cyclostrophic balance between the pressure gradient and the centrifugal forces leads to the collapse of these vorticity clusters into narrow high-energy filaments.

However, in two dimensions, the concept of vortex stretching is nonexistent. If a quasi-two-dimensional model is used to model the behavior of cusps in a shallow boundary layer, the cusps will eventually be subject to stretching, which can then lead to the storm intensification and a tornado-like vortex development. As stretching occurs, the cusps unravel (see Fig.~\ref{fig:coherent}), and the quasi-two-dimensional turbulence model has to be replaced by a three-dimensional model.

The numerical simulations in \cite{markowski24} show that tornado development might be enhanced by turbulence and the dynamics of vortical structures in the boundary layer. Specifically, it is argued that boundary layer ``coherent structures,'' such as cusps and hairpins \cite{adrian07, bernard11, bernard19, markowski24}, play a significant role in tornadogenesis. The initial formation of cusps and hairpins occurs in a relatively thin boundary layer, whose thickness is estimated to be $\approx 6$ meters \cite{markowski24}. When the height of the boundary layer is relatively small (see more on that in Section~\ref{sec:statmech}), we may view the filament model in \cite{kleinmajda91,kleinmajda912} and described in Section~\ref{sec:schrodinger} as representing two-dimensional turbulence. We point out the similarity to the behavior of point vortices in two dimensions, where vorticity evolution can also be governed by other forms of the nonlinear Schr\"odinger equation. A recent work on verification of such analogies, particularly an analogy with the Gross--Pitaevskii equation, is presented in \cite{muller24} and discussed further in Section~\ref{sec:Gross--Pitaevskiis}. A deep mathematical analysis connecting fluid (Euler and Navier--Stokes) equations and various forms of the nonlinear Schr\"odinger equation is presented in \cite{khesin18}.

The results of the numerical simulations in \cite{markowski24} show alternating streaks of wavelength $500$-$700$ m of positive and negative vertical component of vorticity, $\zeta$, on the warm side of the gust front (see Fig.~5 in \cite{markowski24}). In the positive-$\zeta$ streaks, $\zeta$ frequently reaches or exceeds $0.03$ s$^{-1}$ at elevation $7.5$ m. The $\zeta$ in the negative-$\zeta$ streaks is generally of lesser magnitude, rarely less than $-0.02$ s$^{-1}$. This compares favorably to the furrows and hairpin legs in \cite{bernard11} (see Figs.~\ref{fig:bernard1} and \ref{fig:bernard2}) and is reminiscent of the initialization of the surface vertical vorticity in \cite{parker23} (see Fig.~5 in \cite{parker23}). Fig.~7 in \cite{markowski24} shows streamwise oriented $\lambda_2$-isosurfaces consistent with the hairpin legs in \cite{bernard11} and shown in Figs.~\ref{fig:bernard1} and \ref{fig:bernard2}. Similarly, the vortex lines shown in Fig.~\ref{fig:coherent} compare favorably to the solutions of the Schr\"odinger equation shown in Fig.~\ref{fig:schrodgr}. We note that to generate the alternating vorticity streaks in \cite{markowski24}, small random initial temperature perturbations were added and the boundary layer was given $12$ hours to evolve to a quasi-steady state before storms were initiated via the introduction of a warm bubble. It may be interesting to explore the connection between this initialization process and the computational stage in \cite{bernard11} after the flow encounters the thin horizontal plate and before the onset of turbulence (see Figs.~1 and 2 of \cite{bernard11} for $0.5<x<0.8$).

In the video ``Mays's Fury'' \cite{maysfury}, which documents the coverage of a violent tornado outbreak on May 3rd, 1999, a nearly horizontal vortex in or near a tornado while the tornado crosses  interstate I-35 is shown in the 41st minute. In addition, the horizontal vortex is visible in the region where the tornado appears to be the most intense. This is suggestive of a hairpin leg in or adjacent to the tornado and the associated unraveled narrow cusp (vertical filament) vortices, moving into the tornado, causing the tornado to intensify or maintain its intensity.

The paper is organized as follows. In Section~\ref{sec:schrodinger} we discuss the nonlinear Schr\"odinger equations relevant for modeling the behavior of single vortex filaments that capture the self-induction and eventually also stretching consistent with the Navier--Stokes equations. We demonstrate that these models have the ability to model the creation of vortex cusps and are thus relevant for the early stages of tornadogenesis. In Section~\ref{sec:statmech} we review the statistical mechanics of the relevant vortex gas models that allow us to better understand the quasi-two-dimensional turbulence in a shallow boundary layer. We connect these ideas to Stages 2 and 3 of the tornadogenesis process of \cite{fischer24}. Section~\ref{sec:thermodynamics} outlines the foundations of non-equilibrium thermodynamics, addresses the ideas of entropic balance in a tornado environment, and introduces the analysis of macroscopic fluctuations of thermodynamic fluxes in the local entropic balance equation.  We use non-equilibrium thermodynamics to derive an expression for the entropy gradient associated with non-equilibrium turbulent heat flux that, after development of quasi-two-dimensional turbulence, combined with updrafts below the supercell wall cloud, will tilt and stretch the vortex lines and unravel their cusps leading to tornadogenesis. We also provide the expressions for the heat and the entropy production in the boundary layer. In Section~\ref{sec:Gross--Pitaevskiis} we use the analogy between the filament equation and its counterpart in the nonlinear Schr\"{o}dinger equation and develop it further to compare with the (mathematical but not modeling) structure of the special case of the Gross--Pitaevskii equation. We also provide references there on the equivalence of the Schr\"{o}dinger and the Gross--Pitaevskii wave functions and fluid flow equations via the Madelung and Hasimoto transforms. The benefits in this case come from the fact that vortex equations use more fundamental, potential-like quantities, such as the vortex density and the curvature, rather than the original fluid flow fields (velocity and vorticity), to describe flow singularities.

Finally, Section~\ref{sec:summary} provides a summary and discussion of the main results and concludes with ideas for future investigation.

\section{Asymptotic Equation for Vortex Filaments}
\label{sec:schrodinger}

In this section we address the behavior of single vortex filaments that could be in an originally straight, horizontal position, similar to the vortex filaments prior to the onset of cusps and furrows in \cite{bernard11} and discussed in the introduction. One advantage of the approach in this section is that flow singularities can be described by focusing on properties of the vortex, in this case its curvature, rather than on the underlying velocity and vorticity fields that could suffer blow-ups that could be difficult to model~\cite{elgindi21}. Additionally, we will see that vertical cusp-like structures can develop, tying the results of this section to Stage 2 of tornadogenesis discussed in the introduction.

Small-amplitude, short-wavelength perturbations of a single thin horizontal vortex filament are investigated in \cite{kleinmajda91,kleinmajda912} (see also \cite{majdabertozzi02}). This work relates to our understanding of the behavior of the initially horizontal transverse vortices in~\cite{bernard11} and captures the initial stages of the development of cusps and mushroom caps. Similar to \cite{bernard11}, the background flow is a shear flow. In the following paragraphs, we will review the evolution of such a three-dimensional vortex filament using a model that does not allow for stretching and also show how cusps can form. Later in this section, we discuss the extension to a model that allows for stretching, also derived in \cite{kleinmajda91,kleinmajda912}, and its implications.

Consider a narrow, three-dimensional vortex filament described by its center curve position function ${\bf x} (\gamma,t)$, where $\gamma$ is the curve parameter and $t$ is the (rescaled) time variable. Let $\kappa (\gamma,t)$ be the curvature of the center curve and ${\bf n} (\gamma,t)$ the binormal vector along the center curve of the filament (i.e., the cross product of the unit tangent and the unit normal vectors to the curve). The {\it self-induction} equation governing the evolution of the filament and consistent with the Navier--Stokes equations is then \cite{kleinmajda91,kleinmajda912}
\begin{equation}
    \frac{\partial {\bf x}}{\partial t}  = \kappa (\gamma,t) {\bf n} (\gamma,t).
    \label{filament1}
\end{equation}
Note that we use $\bf n$ for the binormal vector instead of the standard $\bf b$ to avoid confusion with the thermodynamic fluxes to be defined and used in a later section. The validity of the self-induction equation \eqref{filament1} rests on the following four assumptions, in which $\Gamma$ is the circulation around the filament, $\nu$ is the kinematic viscosity, ${\boldsymbol\omega}({\bf x},t)$ is the vorticity, and $r(\gamma,t)=\kappa(\gamma,t)^{-1}$ is the radius of curvature of the center curve:
\begin{itemize}
    \item the vorticity ${\boldsymbol\omega} ({\bf x}, t)$ is nonzero only inside the filament of radius $\delta \ll1$;
    \item the radius satisfies $\delta = {(\Gamma/\nu)}^{-1/2}$;
    \item the vorticity ${\boldsymbol\omega} ({\bf x}, t)$ is large enough inside the tube so that $\Gamma/\nu\to\infty$ as $\nu\to0$;
    \item the radius of curvature $r(\gamma,t)$ is bounded strictly away from zero.
 \end{itemize}
We note that equation \eqref{filament1} does not allow for changes of the length of the curve, and thus the dynamics described by \eqref{filament1} will not capture stretching effects \cite{majdabertozzi02}.

\begin{figure}
    \begin{center}
        \includegraphics[width=0.65\textwidth]{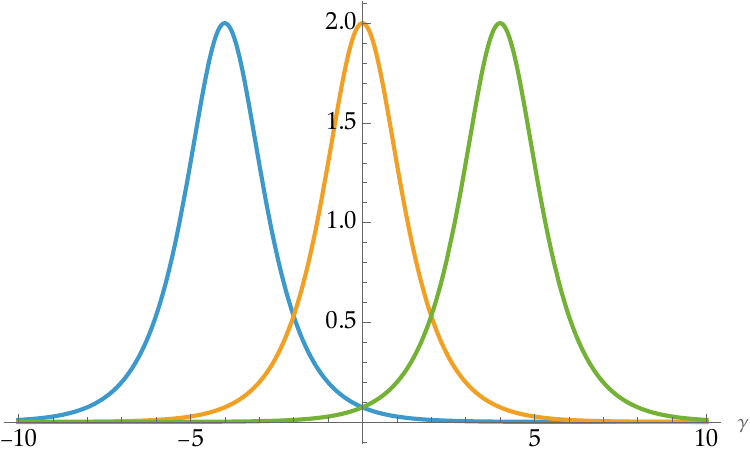}
    \end{center}
    \caption{A visualization of the exact solution \eqref{shr3exact} of the cubic nonlinear Schro\"{o}dinger equation \eqref{shr1} as a function of $\gamma$. The parameter values are $A=B=C_1=1$ and $C_2=0$, and three different times, $t=-2$, $0$, and $2$, are shown left to right.}
    \label{fig:schrodgr}
\end{figure}

Consider next the {\it Hasimoto transform}
\begin{equation}
    \psi (\gamma,t) = \kappa (\gamma,t) e^{i \Phi (\gamma,t)},
    \qquad
    \Phi (\gamma,t) = \int_{0}^{\gamma} \tau(y, t) \, dy,
    \label{hasimoto}
\end{equation}
where $\psi(\gamma,t)$ is the complex-valued {\it filament} function and $\tau(\gamma,t)$ is the torsion of the filament curve \cite{kleinmajda91},\cite{majdabertozzi02}. Recall that torsion can be viewed as a measure of the failure of a curve to be planar. A perturbation of \eqref{filament1} in the form
\begin{equation}
    \label{filament2}
    \frac{\partial {\bf x}}{\partial t}
    =
    \kappa (\gamma,t) {\bf n} (\gamma,t)
    +
    \tilde\delta {\bf v} (\gamma,t),
\end{equation}
where ${\bf v} (\gamma,t)$ is the center curve velocity field, together with the Hasimoto transform \eqref{hasimoto}, then leads to the perturbed Schr\"{o}dinger equation for the filament function $\psi$:
\begin{equation}
    \label{shrpert}
    \frac{1}{i} \frac{\partial \psi}{\partial t}
    =
    \frac{\partial^2 \psi}{\partial \gamma^2}
    +
    \frac{1}{2} |\psi|^2 \psi
    -
    \tilde\delta A
    \left(
        \psi, \frac{\partial {\bf v}}{\partial \gamma}
    \right).
\end{equation}
The expression for the last term in \eqref{shrpert} is given by formula (7.21) in \cite{majdabertozzi02}, and since in this section we will be interested in the case $\tilde\delta=0$, we won't repeat it here. For the original self-induction case with $\tilde\delta=0$, equation \eqref{shrpert} turns into the nonlinear Schr\"odinger equation (with cubic nonlinearity)
\begin{equation}
    \label{shr1}
   \frac{1}{i} \frac{\partial \psi}{\partial t} = \frac{\partial^2 \psi}{\partial \gamma^2} + \frac{1}{2} |\psi|^2 \psi.
\end{equation}
Note from \eqref{hasimoto} that the magnitude of the filament function appearing in \eqref{shrpert} and \eqref{shr1} is the curvature of the filament, i.e.,
\begin{equation*}
    |\psi(\gamma,t)|=\kappa(\gamma,t).
\end{equation*}
The cubic Schr\"odinger equation \eqref{shr1} has solutions (including explicit ones) that allow for large values of $|\psi|$ and thus large curvatures. An example of an explicit solution exhibiting dependence on both $\gamma$ and $t$ is \cite{novikov84}
\begin{equation}
    \psi(\gamma,t)
    =
    \pm2Ae^{i\left(B\gamma+(A^2-B^2)t+C_1\right)}\sech{(A\gamma-2ABt+C_2)}
    \label{shr3exact}
\end{equation}
with arbitrary real constants $A$, $B$, $C_1$, and $C_2$. For this solution we have $\max|\psi(\gamma,t)|=2|A|$, showing that the curvature can be made arbitrarily large. Note that \eqref{shr3exact} imitates a cusp-like and a soliton-like solution, and the visualization of $|\psi(\gamma,t)|$ for fixed values of the constants and three different time values is shown in Fig.~\ref{fig:schrodgr}.
This solution indicates that the cusp structures appearing in the boundary layer and observed in the simulations (see, e.g., \cite{adrian07, bernard11, bernard19, markowski24}) may appear as a result of turbulence in the boundary layer before stretching occurs.

An asymptotic extension of the nonlinear Schr\"odinger filament equation~\eqref{shr1} is derived in \cite{kleinmajda91,kleinmajda912,majdabertozzi02} to allow for stretching. An originally straight filament in the unit direction ${\bf t}_0$ and of core size $\delta$ is replaced with a perturbed version
\begin{equation*}
    {\bf x}(\gamma, t) = \gamma {\bf t}_0 + \varepsilon^2 {\bf X}
    \left(\frac{\gamma}{\varepsilon}, \frac{t}{\varepsilon^2}\right)
    + \smallO(\varepsilon^2),
    \quad
    \varepsilon\ll1,
\end{equation*}
where ${\bf X}\left(\frac{\gamma}{\varepsilon}, \frac{t}{\varepsilon^2}\right)$ is a perturbation term orthogonal to ${\bf t}_0$ and rapidly decreasing in the first variable. For additional assumptions for this model see \cite{kleinmajda91}. This ansatz then leads to an extension of the filament self-induction equation~\eqref{filament1} that has the form
\begin{equation*}
    \frac{\partial {\bf x}}{\partial t}  = \kappa (\gamma,t) {\bf n} (\gamma,t) + \varepsilon^2  I[{\bf X}(\gamma, t)]\times {\bf t}_0,
\end{equation*}
which, recalling the Hasimoto transform \eqref{hasimoto},
\begin{equation*}
    \psi (\gamma,t) = \kappa (\gamma,t) e^{i \Phi (\gamma,t)},
    \qquad
    \Phi (\gamma,t) = \int_{0}^{\gamma} \tau(y, t) \, dy,
\end{equation*}
and assuming $\delta\ll\varepsilon\ll1$ leads to an asymptotic filament equation with self-stretching
\begin{equation}
    \label{shr2}
    \frac{1}{i} \frac{\partial \psi}{\partial t} = \frac{\partial^2 \psi}{\partial \gamma^2} + \varepsilon^2 \left( \frac{1}{2} |\psi|^2 \psi - I[\psi] \right).
\end{equation}

In the above equations, the nonlocal linear operator $I[.]$ is defined as
\begin{equation}
        I[\psi(\gamma, t)]
        \equiv \int_{-\infty}^{\infty}
        \frac{1}{|h|^3}
        \left[
        \psi(\gamma + h,t) - \psi(\gamma,t)
        - h\,\frac{\partial \psi}{\partial \gamma} (\gamma + h,t)
        +\frac{1}{2} h^2 H (1 - |h|)
        \frac{\partial^2 \psi}{\partial \gamma^2}(\gamma,t)
        \right]
        \, dh,
    \label{Iterm}
\end{equation}
where $H$ is the Heaviside function. The operator $I[.]$ is nonlocal in the sense that it includes information from all the points along the centerline of the filament. Note that in \eqref{shr2} the cubic term $\dfrac12|\psi|^2\psi$ and the $I[\psi]$ term compete on the same scale.

\begin{figure}
    \begin{center}
        \includegraphics[width=0.75\textwidth]{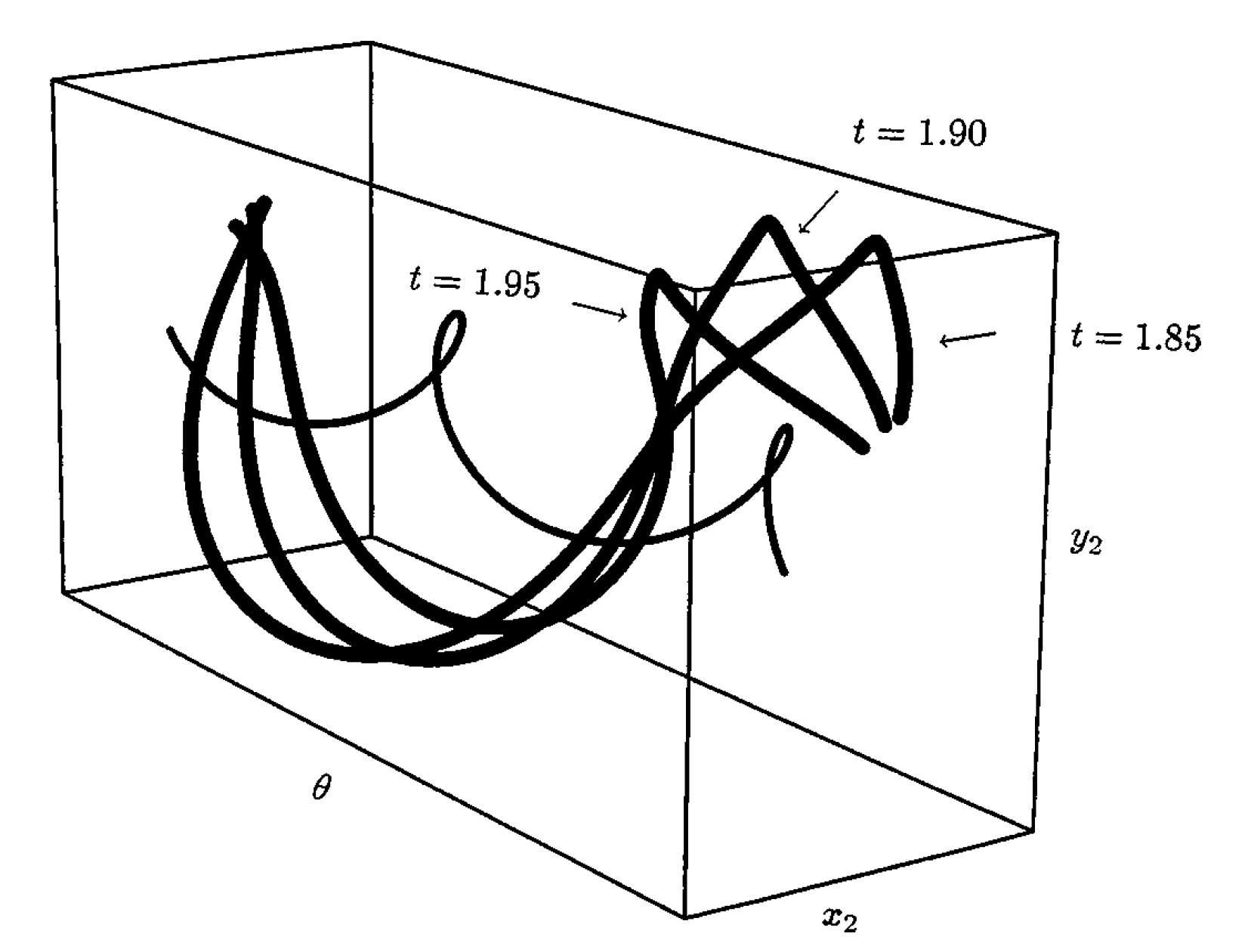}
    \end{center}
    \caption{Time sequence of the filament curve showing the initial helical perturbation of a straight filament at time $t=0$ and three later stages at $t=1.85$, $1.90$, and $1.95$, showing the development of cusps along the filament \cite{kleinmajda912}.}
    \label{fig:filament}
\end{figure}

Equation~\eqref{shr2} models the ``birth'' of strong cusps in the vortex filament from the interaction of the nonlocal operator (representing a strain flow) and the cubic nonlinearity, although the assumptions of the asymptotic theory become invalid as the cusp vortex is born. Specifically, the assumptions of small-amplitude, short-wavelength perturbations of a thin straight vortex are violated just before cusps or any shapes resembling mushroom caps evolve. In the numerical calculations reported in \cite{kleinmajda912} the vortex filament develops higher and much narrower peaks as time evolves when compared with the corresponding solutions of the cubic nonlinear Schr\"odinger equation \eqref{shr1}. The details of the evolution up to that time are fully discussed in \cite{kleinmajda912}. The time sequence of the behavior of an initially helically perturbed straight filament was computed in \cite{kleinmajda912} and is shown in Fig.~\ref{fig:filament}.
The initially helical filament is shown through the middle of the figure, and three stages in the initial development of cusps along the filament are shown at times $t=1.85$, $1.90$, and $1.95$, showing the formation of a cusp near where the arrows in Fig.~\ref{fig:filament} are pointing.

Vortex filaments with vertical segments are important for the formation of vortex patches in Stage 2 of our understanding of tornadogenesis discussed in the introduction. Stage 3 starts with the organization of vortex patches into a single symmetric vortex that later intensifies via stretching. The following section reviews the ideas from statistical mechanics of vortex gases and ties them to the aforementioned organization and single vortex generation.

\section{Statistical Mechanics of Vortex Gases}
\label{sec:statmech}
In statistical mechanics of gases, solving for the behavior of a large number of particles (gas molecules) is simplified by making assumptions about the mean behavior of the collection, rather than attempting to solve the intractable system of equations for the interactions of each particle with all the others. The phase space of such a system contains information about the position and momentum of each particle, and assumptions are made about the energy, $E$, and entropy, $S$, of the system, leading to a probability distribution on the phase space. In the case of fluid dynamics, the phase space of vortex gases depends on the set up of the vortex gas, which could be point vortices in 2D (see, e.g., sections 6.2 and 7.3 of \cite{majdabertozzi02}, parallel or nearly parallel vortices in 3D (section 7.4 of \cite{majdabertozzi02}), or a single vortex filament \cite{belikdokkenpotvinscholzshvartsman17,belikdahldokkenpotvinscholzshvartsman18,belikdokken23}. Parallel vertical vortices are analogous to a corresponding set of point vortices in the plane in that they satisfy the same equations \cite{majdabertozzi02}.

In the case of an ideal gas, let the microstates be indexed by $i$. We denote the energy of microstate $i$ by $E_i$ and the partition function corresponding to this scenario is
\begin{equation}
    Z=\sum_i e^{-\beta E_i},
    \label{eq:partition}
\end{equation}
where $\beta=\partial S/\partial E$ represents the inverse temperature, i.e., $\beta=1/T$. A similar approach, possibly with the sum in the partition function \eqref{eq:partition} replaced by an integral of $e^{-\beta E}$ over all configurations of vortices under consideration, can be used in fluid dynamics.

From probabilistic considerations, when $\beta=0$ (or $T=\pm\infty$), all energy levels are equally likely to occur, and the entropy $S$ is at its highest point. For $\beta>0$, configurations with lower energy are more likely and these correspond to more balled up vortices. Finally, for $\beta<0$, configurations with higher energy are more likely and these correspond to straighter vortices. See \cite{belikdokkenpotvinscholzshvartsman17,belikdahldokkenpotvinscholzshvartsman18,belikdokken23} for details. Therefore, negative temperature vortices are hotter than positive temperature vortices, with vortices with $T=\pm\infty$ ($\beta=0$) temperatures in between them.

Using a different viewpoint, we can say that negative temperature vortices have higher energy and are (more) straight or smooth, while positive temperature vortices are (more) balled up and, from a physical point of view, dissipating. Therefore, energy is expected to flow in the direction of increasing $\beta$: straight, high energy vortices will become less straight and begin to fold up as $\beta<0$ increases towards $0$. Such vortices will transfer their energy to larger scales. As $\beta$ increases into positive values, the configurations become more balled up, starting the Kolmogorov energy cascade to smaller scales and dissipating \cite{belikdokkenpotvinscholzshvartsman17,belikdahldokkenpotvinscholzshvartsman18,belikdokken23,chorin,newton01}. We note that for negative temperatures to occur, the phase space must be bounded, as well as the energies $E_i$ \cite{newton01}.

Let's next consider a collection of nearly parallel vortices considered in \cite{kleinmajdadamodaran95} with their principal direction ${\bf t}_0$ assumed to be vertical. This approach can be used to model the various nearly vertical vortices in a shallow layer. For future considerations we also assume this shallow layer is above a bounded region on the surface plane beneath a supercell. Similar to the description in Section~\ref{sec:schrodinger}, the general center curve of each of the filaments, ${\bf x}_i$, is assumed to have the asymptotic form
\begin{equation*}
    {\bf x}_i(\gamma, t) = \gamma {\bf t}_0 + \varepsilon^2 {\bf X}_i
    \left(\frac{\gamma}{\varepsilon}, \frac{t}{\varepsilon^4}\right)
    + \smallO(\varepsilon^2),
    \quad
    \varepsilon\ll1,
\end{equation*}
where ${\bf X}_i\left(\frac{\gamma}{\varepsilon}, \frac{t}{\varepsilon^4}\right)$ is again a perturbation term orthogonal to ${\bf t}_0$. For additional assumptions for this model, similar to those in Section~\ref{sec:schrodinger}, see \cite{kleinmajdadamodaran95}. We point out that the separation distance between filaments has to be much larger than the filament core size. It is then possible to derive asymptotic limiting equations for $N$ nearly parallel interacting vortex filaments
\begin{equation}
    \frac{\partial{\bf x}_i}{\partial t}
    =
    J\left[
    \alpha_i\Gamma_i\frac{\partial^2{\bf x}_i}{\partial\gamma^2}
    \right]
    +
    J\left[
    \sum_{j\ne i}^N
    2\Gamma_j\frac{{\bf x}_i-{\bf x}_j}{|{\bf x}_i-{\bf x}_j|^2}
    \right],
    \quad
    1\le i\le N,
    \label{eq:kmd95}
\end{equation}
where $J$ is the skew-symmetric matrix
\begin{equation*}
    J=\begin{pmatrix}
        0 & -1 \\
        1 & 0
    \end{pmatrix}.
\end{equation*}
Here, $\Gamma_i$ is the circulation of the $i$th vortex and $\alpha_i$ is a constant determined by the vortex core structure (see additional references in \cite{kleinmajdadamodaran95}). Using the identification $\psi_j(\gamma,t)=x_j(\gamma,t)+iy_j(\gamma,t)$, where ${\bf x}_j=(x_j,y_j,\gamma)$, equations~\eqref{eq:kmd95} can be re-written as coupled nonlinear Schr\"odinger equations. A special example illustrating their structure is for a pair of two interacting vortices with identical core parameters $\alpha_1=\alpha_2=1$ and circulations $\Gamma_1=1$ and $\Gamma_2=\Gamma$. In that case, making a change of variables $\psi=\psi_1-\psi_2$ and $\phi=\psi_1+\psi_2$, system~\eqref{eq:kmd95} becomes
\begin{equation}
    \begin{aligned}
        \frac{1}{i}\frac{\partial\phi}{\partial t}
        &=
        \frac12(1+\Gamma)\frac{\partial^2\phi}{\partial\gamma^2}
        +
        \frac12(1-\Gamma)
        \left[
            \frac{\partial^2\psi}{\partial\gamma^2}-4\frac{\psi}{|\psi|^2}
        \right],\\
        \frac{1}{i}\frac{\partial\psi}{\partial t}
        &=
        \frac12(1-\Gamma)\frac{\partial^2\psi}{\partial\gamma^2}
        -
        \frac12(1+\Gamma)
        \left[
            \frac{\partial^2\psi}{\partial\gamma^2}-4\frac{\psi}{|\psi|^2}
        \right],
    \end{aligned}
    \label{eq:2vortices}
\end{equation}
which interestingly becomes decoupled for the case of co-rotating filaments of equal strength ($\Gamma=1$). Numerical evidence is presented in \cite{kleinmajdadamodaran95} for a finite-time collapse of a vortex pair with a negative circulation ratio $\Gamma$ in the sense that the filaments would eventually come close to each other and violate the assumption on the separation distance between the vortices. On the other hand, numerical evidence is given that no finite-time collapse happens for a positive circulation ratio. Notice in Figs.~\ref{fig:bernard1} and \ref{fig:bernard2} how the legs of the cusps, originally at a non-zero distance from each other at times $t<0.7$ appear to have merged after time $t=0.8$, consistent with the first scenario. Also notice that the nonlinearity in \eqref{eq:2vortices} is not cubic, making these equations qualitatively different from those for a single vortex filament (equations \eqref{shr1} and \eqref{shr2}).

The shallow boundary layer with cusps below the supercell has filaments in nearly vertical directions and therefore the nearly parallel vortex gas model should be applicable. A further simplification and insight come from assuming that the vortex filaments are actually vertical and parallel to each other. In that case, the second-derivative with respect to $\gamma$ terms in \eqref{eq:kmd95} and \eqref{eq:2vortices} are equal to $0$, and the equations become identical to those for point vortices in two dimensions (see section 7.3 of \cite{majdabertozzi02}). Hence, the term {\it quasi-two-dimensional turbulence} is used to describe three-dimensional turbulence in a shallow surface layer when the vortices can be approximated as vertical and straight. In view of the earlier discussion, such vortices, being perfectly straight, can be considered to have negative statistical mechanical temperature.

We now review the results of the two-dimensional model and connect them to the formation of the vortex patches. The two-dimensional mean field theory was developed in \cite{caglioti91,caglioti95} and allows for a range of negative temperatures. This theory motivated a three-dimensional theory \cite{lions00}, which does not include negative temperatures; however, an extension of this work allows for a small range of negative temperatures provided the 3D vortex has a fractal cross section \cite{flandoli02}.

An $N$-vortex system in a bounded domain $\Lambda$ in $\mathbb{R}^2$ with the associated canonical Gibbs measure at inverse temperature $\tilde \beta$ is studied in \cite{caglioti91,caglioti95}. The vortices are all assumed to have the same intensity $\alpha$ (we are using the notation of \cite{caglioti91}, not to be confused with the 3D vortex core parameters $\alpha_i$ earlier in this section). It is shown that as $N\to\infty$, $\tilde\beta/N\to\beta\in(-8\pi,+\infty)$, and $\alpha N\to1$, the stream function $\psi$ associated with this problem satisfies the mean field equation
\begin{equation}
        -\Delta\psi
        =
        \frac{e^{-\beta\psi}}{\int_\Lambda e^{-\beta\psi}},
        \qquad
        \psi
        =
        0
        \quad
        \text{on }\partial\Lambda.
        \label{eq:mfe}
\end{equation}
The limiting vorticity field then satisfies $\boldsymbol\omega_\beta=-\Delta\psi\ge0$ and $\int_\Lambda\boldsymbol\omega_\beta=1$. For a disk domain $\Lambda$, the solutions of the mean field equation \eqref{eq:mfe} can be found and satisfy $\boldsymbol\omega_\beta\to\delta_{x_0}$ as $\beta \to-{8\pi}^+$, where $x_0$ is the center of the disk and $\delta_{x_0}(x)$ is the Dirac delta function. It can also be shown that this kind of vorticity concentration also exists in a simply connected domain sufficiently close to a disk and possibly also in other domains. However, there are domains for which this concentration does not occur (e.g., rectangles with large aspect ratios) \cite{caglioti95}. Convergence to a smooth solution is also possible \cite{caglioti91,caglioti95}. If we were to consider the extension of this 2D model back to the case of co-rotating vertical parallel vortex filaments, the scenario in which vorticity tends to concentrate near a single point can be connected to the organization of vertical vorticity into a single symmetric intense vertical vortex that would later undergo stretching and intensify further.

In the case of the formation of furrow-like structures with cusps and adjacent hairpin legs present in the numerical simulations \cite{bernard11, markowski24}, we let $\Lambda$ to be a narrow, bounded region with a smooth boundary below the furrow between the hairpin legs. In the presence of many furrows, each such vortex patch would be modeled by a corresponding region $\Lambda$. Figs.~\ref{fig:bernard2} and \ref{fig:coherent} showing the vertical vorticity cusps suggest that the total vorticity in the quasi-two-dimensional turbulence patches (regions $\Lambda$) should be zero, provided the positive and negative vorticity legs have the same strength. However, a 12-hour burn-in that included Coriolis effects (planetary vorticity) was used in the simulations of \cite{markowski24}, which led to the positive vorticity leg having a larger vorticity value than the negative vorticity leg, or, in the language of \cite{markowski24}, the positive-$\zeta$ anomalies in the boundary layer were stronger than the negative-$\zeta$ anomalies. Consequently, the total vertical vorticity induced by the cusps in the patches is expected to be positive.

If we think of the vertical vorticity cusps as straight vertical filaments and the turbulent boundary layer as made up of these filaments, we can use the two-dimensional turbulence model of \cite{caglioti91, caglioti95} discussed above. With the filaments being straight, we can assume that they have negative temperature and cluster into larger filaments with cyclonic and anti-cyclonic legs (the right image of Fig.~\ref{fig:bernard2}). As these larger clusters in the quasi-two-dimensional turbulence are tugged up into the mesocyclone (in a streamwise sense), the negative side of the vertical vorticity cusp tilts $180$ degrees, aligns with the positive side, and the vertical vorticity component becomes positive (see Fig.~\ref{fig:coherent}). As this happens and the vortex stretches, a transition from a quasi-two-dimensional model to a three-dimensional model is needed.

Initially, the resulting vortex is in cyclostrophic balance, i.e., the pressure gradient force and the centrifugal force are in balance. Surface friction along with the stretching and rotation of the vortex causes the centrifugal force to weaken and allows the pressure gradient force to dominate. Consequently, the vortex collapses to a narrow filament with high energy density. This organization in the flow results in a decrease in entropy density, increase in energy density, and a negative temperature of the vortex since $\Delta S/\Delta E<0$ \cite{belikdokkenpotvinscholzshvartsman17,belikdahldokkenpotvinscholzshvartsman18,belikdokken23}. Such vortices are called supercritical or suction vortices and can be approximated by a delta function $\delta_{x_0}(x)$ supported on a vertical segment. As discussed above, these supercritical vortices may then contribute to tornadogenesis. Further stretching of a negative temperature vortex would lead to its transferring energy to larger scales via folding, and eventual dissipation of energy via the Kolmogorov cascade, a direct energy cascade to smaller and smaller scales without lower bound \cite{belikdokkenpotvinscholzshvartsman17,belikdahldokkenpotvinscholzshvartsman18,belikdokken23}. The processes described in the above may be occurring repeatedly in the region below the supercell and mesocyclone.

\section{Non-Equilibrium Thermodynamics and Entropy}
\label{sec:thermodynamics}

In this section we outline the foundations of non-equilibrium thermodynamics and address the entropic balance under the highly non-equilibrium settings. In subsection~\ref{sec:entropy} we list the expressions for entropy flux, the heat flux, and the internal entropy production in classical irreversible and extended irreversible thermodynamics. We also discuss entropic balance in the boundary layer that is instrumental in stretching and tilting of vorticity into the vertical direction, which is a part of Stages 3 and 4 of tornadogenesis described in the introduction. We provide the entropy balance equations in classical and extended forms including the internal entropy production term. In subsection~\ref{sec:meso} we employ the {\it macroscopic fluctuations} analysis to show that the entropy flux and must be a function of entropy. We also provide a connection between entropy gradient with local vorticity and discuss non-equilibrium thermodynamics of the boundary layer including the heat and the entropy production in the boundary layer.

We should note that in recent years studies of ``standard'' classical thermodynamics have developed in many branches influenced by non-equilibrium models in phenomenological and statistical formulations. In this paper, we are carefully specifying the thermodynamical context we use. For instance, this section applies ``traditional'' irreversible thermodynamics in (subsection~\ref{sec:entropy}) and the recent developments in much less traditional theory of macroscopic fluctuations (subsection~\ref{sec:meso}). However, the preceding Section~\ref{sec:statmech} uses thermodynamics and statistical mechanics of vortex gases. We also emphasize that these two approaches are independent of one another even though there are some formal similarities.

\subsection{Entropy in Irreversible Non-Equilibrium Thermodynamics}
\label{sec:entropy}
This subsection follows the standard development of the irreversible thermodynamic theory \cite{lebon08}. We begin with {\it classical irreversible thermodynamics}. The first step to modeling a system that is not in equilibrium is to employ the {\it local equilibrium hypothesis}, where all the relevant thermodynamic variables are defined locally and are in a quasi-equilibrium for a ``small enough'' system. In particular, let $s({\bf x},t)$ denote the specific entropy as a function of position, ${\bf x}$, and time, $t$, and assume also that it is a function of the extensive specific variables $a_j({\bf x},t)$, whose corresponding conjugate intensive variables are denoted by $F_j({\bf x},t)$. In the fluid dynamics context, such a ``small enough'' system is an air parcel. The Gibbs' equation
\begin{equation}\label{locDelta}
    D s ({\bf x},t) =
    \sum_{j} F_j ({\bf x},t) \, D a_j ({\bf x},t)
\end{equation}
gives the relationship between these variables and leads to the differential equation
\begin{equation*}
    \frac{D s ({\bf x},t)}{Dt} =
    \sum_{j} F_j ({\bf x},t) \, \frac{D a_j ({\bf x},t)}{Dt},
\end{equation*}
where
\begin{equation*}
    \frac{D}{Dt} =  \frac{\partial}{\partial t}
    + {\bf v} \cdot \nabla
\end{equation*}
is the material derivative, i.e., the rate of the change of a quantity following the trajectory of the air parcel. The balance relation for the specific entropy is then
\begin{equation}
    \label{stransport}
    \rho\,\frac{D s }{Dt} =
     - \nabla \cdot {\bf B}
     + \sigma,
\end{equation}
where ${\bf B}({\bf x},t)$ is the {\it entropy flux}, $\rho({\bf x},t)$ is the mass density, and $\sigma({\bf x},t)\ge0$ is the entropy production rate per unit volume \cite{lebon08, liu08, demirelgerbaud19}. The entropy production rate is nonnegative due to the second law of thermodynamics and, in general, incorporates the following sources \cite{demirelgerbaud19}:
\begin{enumerate}
    \item entropy production associated with heat transfer;
    \item entropy production due to mass transfer;
    \item entropy production because of viscous dissipation of fluid;
    \item entropy production arising from chemical reactions.
\end{enumerate}
For simplicity, our discussion will focus solely on the entropy production associated with heat transfer; however, we acknowledge that it would also be important to consider the entropy production resulting from mass transfer (moisture flux), which will be left for future work. Transport equations similar to \eqref{stransport} are valid for all the extensive state variables $a_j$. We use the basic thermodynamic model with two extensive local state variables, the specific internal energy, $u({\bf x},t)$, and the specific volume, $v({\bf x},t)=\rho^{-1}({\bf x},t)$, so that the equation of state is $s = s(u,v)$. For the purposes of this discussion, we assume that both state variables $u$ and $v$ are known functions of ${\bf x}$ and $t$. Consequently, we have two conjugate intensive variables here,
\begin{equation*}
    \frac{\partial s}{\partial u} = T^{-1}
    \quad \text{ and } \quad
    \frac{\partial s}{\partial v} = pT^{-1},
\end{equation*}
where $T({\bf x},t)$ is the absolute temperature and $p({\bf x},t)$ is the pressure.

In the case of a laminar motion (turbulence is excluded) of a one-constituent isotropic, incompressible ($\nabla\cdot{\bf v}=0$), and viscous fluid in the presence of a temperature gradient, without internal energy supply, we have
\begin{equation}
    \rho\,\frac{Du}{dt}
    =
    -\nabla\cdot{\bf q}
    -
    {\bf P}:{\bf V}
    \quad \text{ and } \quad
    \rho\,\frac{Dv}{dt}
    =
    0,
    \label{eq:balance_eqns}
\end{equation}
where ${\bf q}$ is the heat flux, ${\bf v}$ is the velocity, ${\bf V}=\dfrac12\left(\nabla{\bf v}+\left(\nabla{\bf v}\right)^T\right)$, and ${\bf P}$ is the symmetric pressure tensor, all of these being functions of $\bf x$ and $t$. We recall that ${\bf A}:{\bf B}=\tr\left({\bf A}^T{\bf B}\right)=A_{ij}B_{ij}$ using the Einstein summation notation here. If we decompose the pressure tensor
\begin{equation*}
    {\bf P}=(p+p^v){\bf I}+{\bf P}^0,
\end{equation*}
where $p$ is the hydrostatic pressure and $p^v=\frac13\tr{({\bf P}-p{\bf I})}$ so that ${\bf P}^0$ is traceless, then the balance equations \eqref{eq:balance_eqns} can be substituted into
\begin{equation*}
    \frac{Ds}{Dt}
    =
    \frac{\partial s}{\partial u}\frac{Du}{Dt}
    +
    \frac{\partial s}{\partial v}\frac{Dv}{Dt}
\end{equation*}
to obtain
\begin{equation*}
    \rho\,\frac{Ds}{Dt}
    =
    -\nabla\cdot(T^{-1}{\bf q})
    +
    {\bf q}\cdot\nabla T^{-1}
    -
    T^{-1}\left({\bf P}^0:{\bf V}^0\right),
\end{equation*}
where ${\bf V}^0={\bf V}-\frac13(\tr{\bf V}){\bf I}={\bf V}$ is the traceless part of ${\bf V}$, coinciding with $\bf V$ in the case of an incompressible flow ($\frac13\tr{\bf V}=\nabla\cdot{\bf v}=0$). In the remaining part of this section, we will therefore use ${\bf V}$ in place of ${\bf V}^0$. Comparing this expression with \eqref{stransport} and arguing as in \cite{lebon08} that since $\sigma$ represents the rate of entropy production inside the body, its expression cannot contain a ﬂux term like $\nabla\cdot({\bf B}-T^{-1}{\bf q})$, we obtain the equations
\begin{equation}
    \nabla\cdot{\bf B}({\bf x},t)
    =
    \nabla\cdot\left(T^{-1}({\bf x},t){\bf q}({\bf x},t)\right)
    \label{eq:entropy_flux}
\end{equation}
and
\begin{equation}
    \sigma({\bf x},t)
    =
    {\bf q}({\bf x},t)\cdot\nabla T^{-1}({\bf x},t)
    -
    T^{-1}({\bf x},t)\left({\bf P}^0({\bf x},t):{\bf V}({\bf x},t)\right).
    \label{eq:sigma1}
\end{equation}
Note that \eqref{eq:entropy_flux} can be interpreted as ${\bf B}=T^{-1}{\bf q}+{\bf f}$, where $\nabla\cdot{\bf f}=0$. Since there is no constitutive formula for the internal entropy production, the {\it classical irreversible thermodynamics} postulates the existence of a linear relationship between the thermodynamic fluxes in \eqref{eq:sigma1} and their respective conjugate ``thermodynamic forces.'' Specifically, we have the following phenomenological relationships \cite{lebon08}
\begin{equation}
    {\bf q}
    =
    -\lambda\nabla T
    =
    \lambda T^2\nabla T^{-1}
    \quad\text{ and }\quad
    {\bf P}^0
    =
    -2\eta{\bf V},
    \label{eq:phenom}
\end{equation}
where the first relationship is just the Fourier law with the heat conductivity $\lambda$ and $\eta$ is the dynamic shear viscosity. Using \eqref{eq:phenom}, we can rewrite \eqref{eq:sigma1} as
\begin{equation}
    \sigma({\bf x},t)
    =
    \lambda T^{-2}({\bf x},t)|\nabla T({\bf x},t)|^2
    +
    2\eta T^{-1}({\bf x},t)\left({\bf V}({\bf x},t):{\bf V}({\bf x},t)\right),
    \label{eq:sigma2}
\end{equation}
where $|\cdot|$ stands for a vector norm, and it follows that the entropy transport equation \eqref{stransport} can be written, dropping the explicit dependence on $\bf x$ and $t$, as
\begin{equation*}
    \rho\,\frac{Ds}{Dt}
    =
    \nabla\cdot\left(\lambda T^{-1}\nabla T\right)
    +
    \lambda T^{-2}|\nabla T|^2
    +
    2\eta T^{-1}\left({\bf V}:{\bf V}\right).
\end{equation*}

To go beyond the local equilibrium hypothesis and to extend the relationships like the Fourier law, one can turn to {\it extended irreversible thermodynamics}. In a highly non-equilibrium system, heat flux may not be well approximated by the Fourier law as a non-equilibrium system may have spatial and temporal non-local influence. In this approach, dissipative fluxes are elevated to the status of state variables. If we, for example, extend entropy to depend on the heat flux, i.e., $s=s(u,{\bf q})$, then
\begin{equation}\label{sextendq}
    \frac{D  s} {Dt}
    =
    \frac{\partial s}{\partial u} \frac{D u}{Dt}
    +
    \frac{\partial s }{\partial {\bf q}} \cdot \frac{D {\bf q}} {Dt}
    =
    T^{-1} \frac{D u}{Dt}
    +
    \frac{\partial s }{\partial {\bf q}} \cdot \frac{D {\bf q}} {Dt},
\end{equation}
where we assume that the non-equilibrium temperature is well approximated by the local equilibrium temperature $T$. The other partial derivative is assumed to satisfy
\begin{equation}
    \frac{\partial s}{\partial {\bf q}}
    =
    - T^{-1}\,v\,\alpha (u){\bf q},
    \label{eq:force_sq}
\end{equation}
and the transport equation \eqref{stransport} can then be written as
\begin{equation}
    \rho\,\frac{Ds}{Dt}
    =
    -\nabla\cdot\left(T^{-1}{\bf q}\right)
    +
    {\bf q}\cdot\left(\nabla T^{-1}-T^{-1}\alpha(u)\,\frac{D{\bf q}}{Dt}\right)
    +
    2\eta T^{-1}\left({\bf V}:{\bf V}\right).
    \label{eq:transport_eit}
\end{equation}
Additionally, if we use the Onsager linear assumption between the thermodynamics flux $\dfrac{D{\bf q}}{Dt}$ and its corresponding thermodynamic force in \eqref{eq:force_sq}, we arrive at a Cattaneo-type relation
\begin{equation}
    \frac{D{\bf q}}{Dt}
    =
    -\tau_R^{-1}\left({\bf q}+\lambda\nabla T\right),
    \label{eq:cattaneo}
\end{equation}
where $\tau_R$ denotes the heat flux relaxation time. Equation~\eqref{eq:cattaneo} can be substituted into \eqref{eq:transport_eit} to remove the dependence of the right-hand side on the material derivative of the heat flux.

If the heat flux is concentrated in the boundary layer, where the pressure changes (on average) are insignificant, we can use the Fourier law in \eqref{eq:phenom} and the divergence theorem to estimate the heat supply $Q$ to a region ${\Omega}$ (a thin 3D region with a piecewise smooth boundary) of the quasi-two-dimensional turbulent flow as
\begin{equation*}
    Q
    =
    -\int\int_{\partial\Omega}{\bf q}({\bf x},t)\cdot{\bf n}({\bf x})\,d{\bf x}\,dt
    =
    \int\int_{\Omega}\lambda\Delta T({\bf x},t)\,d{\bf x}\,dt,
\end{equation*}
where ${\bf n}({\bf x})$ is the outer unit normal vector and we assume $\lambda$ is a constant. Using \eqref{eq:entropy_flux} and \eqref{eq:phenom}, the entropy supply to $\Omega$, denoted by $\Delta S_1$, could be calculated as
\begin{align*}
    \Delta S_1
    &=
    -\int\int_{\partial\Omega}{\bf B}({\bf x},t)\cdot{\bf n}({\bf x})\,d{\bf x}\,dt \\
    &=
    -\int\int_\Omega\nabla\cdot{\bf B}({\bf x},t)\,d{\bf x}\,dt \\
    &=
    -\int\int_\Omega\nabla\cdot\left(T^{-1}({\bf x},t){\bf q}({\bf x},t)\right)\,d{\bf x}\,dt \\
    &=
    \int\int_\Omega\nabla\cdot\left(\lambda T^{-1}({\bf x},t)\nabla T({\bf x},t)\right)\,d{\bf x}\,dt \\
    &=
    \int\int_\Omega\lambda\left(\nabla(T^{-1})({\bf x},t)\cdot\nabla T({\bf x},t) + T^{-1}({\bf x},t)\Delta T({\bf x},t)\right)\,d{\bf x}\,dt,
\end{align*}
where the first term in the last expression is due to the heat flux and the second term is due to the heat diffusion. Finally, using \eqref{eq:sigma2}, the entropy production in $\Omega$, denoted by $\Delta S_2$, can be calculated as
\begin{equation}
\begin{aligned}
    \Delta S_2
    &=
    \int\int_\Omega \sigma({\bf x},t)\,d{\bf x}\,dt \\
    &=
    \int\int_\Omega
    \lambda T^{-2}({\bf x},t)|\nabla T({\bf x},t)|^2
    +
    2\eta T^{-1}({\bf x},t)\left({\bf V}({\bf x},t):{\bf V}({\bf x},t)\right)\,d{\bf x}\,dt,
\end{aligned}
    \label{eq:deltaS2}
\end{equation}
which is clearly nonnegative in agreement with the second law of thermodynamics. Notice that equation \eqref{eq:deltaS2} implies that the only flows with zero rate of production of entropy would be those that have constant temperature $T$ and are a solid core rotation about an axis.

A detailed account of recent developments in understanding the irreversible entropy production is given in \cite{land21}. Note that \eqref{eq:transport_eit} gives the evolution of entropy far from the equilibrium and includes contribution due to the turbulent fluxes that, in turn, will affect vorticity development in Stage 3 described in the introduction. In the next section we will study the spatial gradient of the specific entropy, $\nabla s$, and its impact on the tilting, stretching, and amplification of vorticity associated with the quasi-two-dimensional turbulent motion leading to tornadogenesis.

\subsection{Mesoscopic Thermodynamics, Macroscopic Fluctuations, and Vorticity}
\label{sec:meso}
The variations in entropy $s({\bf x},t)$
are responsible for the ``mismatch'' between macroscopic and microscopic descriptions, since they take place on different space and time scales. So, turbulent thermodynamic systems may be categorized as mesoscopic \cite{lebon08, bertini15}. Mismatches due to fluctuations in the velocity field are called {\it turbulent fluxes} \cite{wang20, holton19, dokken06}.  Assuming that conduction is slower than convection and assuming that atmosphere consists of dry air and water vapor, thus neglecting the liquid phase, implies that convection dominates the entropy production \cite{demirelgerbaud19,liu08,hauf87}.

Tornado-like vortices, from radar observations and numerical simulations \cite{orf17, markowski24, belikdokken23} exhibit exactly such behavior of a system far from equilibrium, so they can be treated as  mesoscopic systems; in other words, they exhibit both macroscopic and microscopic features. It would be important then to understand how far from equilibrium that system is. The foundation of such an analysis is to evaluate the probabilities of various states that are far from equilibrium. So, essentially, thermodynamics cannot ignore spatial and temporal ``memory'' as we indicated in the previous subsection~\ref{sec:entropy}.
\footnote
{
Recall that the statistical entropy of a system is defined by $\overline{S} = k_B \ln W$, where $k_B$ is the Boltzmann constant and $W$ is the number of microstates corresponding to the macrostate with entropy $\overline{S}$ which is equivalent to thermodynamic entropy. Then the number of microstates is $W = e^{\overline{S}/k_B}$.
}
This serves as a basis for the Einstein formula that asserts that the probability of the state with entropy $\overline{S}$ can be written as
\begin{equation}
    {\mathbb P} = {\mathbb P}_0 e^{\overline{S}/k_B},\label{einst}
\end{equation}
where ${\mathbb P}_0$ is a normalizing factor \cite{einstein10}. These ideas were developed by many researchers throughout recent decades and have been summarized in \cite{lebon08,bertini15}. The entropy is replaced by a functional that depends on the fluctuations given by the non-equilibrium turbulent flux ${\bf q}({\bf x},t)$. Equation \eqref{einst} then becomes
\begin{equation}
    \label{einst1}
    {\mathbb P} = {\mathbb P}_0 e^{-\overline{Q}/(k_B T)},
\end{equation}
where $\overline{Q}$ is heat delivered to the region $\Omega$ and ${\mathbb P}$ is the probability of the fluctuation of the system caused by that turbulent flux. In \cite{bertini15, eyink90} it is shown that if \eqref{einst1} holds, then the entropy flux ${\bf B}$ must satisfy
\begin{equation}
    {\bf B}(s) = - D(s) \nabla s
    + \chi (s)  {\bf E} (t),
    \label{qflux1}
\end{equation}
where $D(s)$ is the matrix diffusion coefficient, ${\bf E}(t)$ is the convection external field, and where $\chi (s) $ is the mobility matrix, assuming linear response. In many cases $\chi$ is quadratic in $s$ \cite{bertini15}. Equation \eqref{qflux1} implies that the entropy gradient can be expressed as
\begin{equation}\label{grads}
    \nabla s = - D(s)^{-1}\,({\bf B}(s)
    -\chi(s) {\bf E}(t)).
\end{equation}
We note that the convection external field ${\bf E}(t)$ can be measured experimentally \cite{bertini15,zhao21}. The standard approach to connecting the equations of motion and thermodynamics uses buoyancy in the Boussinesq approximation \cite{holton19}. However, the Boussinesq approximation imposes (horizontal) restrictions on density variations. In the general case, the connection between the dynamics of motion and the flow thermodynamics can be seen from the fact that the quantity $v\nabla\theta\cdot{\boldsymbol\omega}$ must remain constant along the parcel trajectories for an isentropic inviscid flow, in which vorticity ${\boldsymbol\omega}$ is tied to the potential temperature $\theta$ \cite{dutton76}. The relationship between different definitions of potential temperature and entropy is given in \cite{hauf87}.

In \cite{sasaki14}, a theory of entropic balance is developed based on variational principles under the assumption of short time scales for phase changes. The governing equations of fluid flow are derived using a variational approach in the thermodynamic context, and a discussion of how entropic balance can be affected by a highly non-equilibrium storm environment is included. In particular, it is stated that the entropic balance must be complemented with a term accounting for non-equilibrium entropy fluctuations, which would be similar to the term described in \eqref{eq:transport_eit}. A comment is included on fluctuations not being averaged on all scales, thus leading to a mesoscopic model of the non-equilibrium system (see equation (5.6b) in \cite{sasaki14}).

More specifically, it is hypothesized that the ``phase-change time scale is significantly shorter than the time scales of convective storms and tornadoes (hypothesis 1)'' and that ``variations of the initial entropy levels are small enough and allow us to approximate them by their ensemble means (hypothesis 2).'' The second of these is the local equilibrium hypothesis discussed in subsection~\ref{sec:entropy}. Consequently, this entropic balance theory is a special detailed case of non-equilibrium thermodynamics and thus is fundamental in our study of the boundary layer. In~\cite{sasaki14}, regions are identified as entropic sources and entropic sinks within the entire supercell storm. In our setting, we consider entropic sources and entropic sinks in the context of the boundary layer. The entropic sink is the evaporatively cooled boundary layer in the forward flank or rear flank of the supercell. The entropic source is the boundary layer, laden with moisture from the northward advection of water vapor from the Gulf of Mexico.

The region of the two indexed flows can be thought of as a box below the wall cloud, just on the cool side of the forward-flank downdraft. In that region the laws of non-equilibrium thermodynamics apply. We assume that the construction in \cite{sasaki14} is applicable in the region of the forward-flank downdraft, not the entire supercell. The close proximity of the entropic source and sink determines the size of the gradient of the entropy. Using a variational argument, entropy considerations, and the Clebsh transformation, it follows that
\begin{equation}\label{clebsh}
    {\bf v} = - \nabla A - s \nabla R
    = - \nabla A + {\bf v}_R,
    \qquad
    {\boldsymbol\omega} = - \nabla s \times \nabla R,
\end{equation}
where $A$ and $R$ are the Lagrange multipliers corresponding to conservation of mass and entropy, respectively, and ${\bf v}_R$ is the rotational component of velocity. We note that the decomposition in \eqref{clebsh} is not unique, but it significantly complements the discussion of \cite{dutton76} on the nexus between vorticity and thermodynamics. By rewriting the second equation in \eqref{clebsh} as
\begin{equation}\label{clebshomega}
    {\boldsymbol\omega}
    = - \frac{\nabla s}{s} \times s\nabla R
    = \frac{\nabla s}{s} \times {\bf v}_R,
\end{equation}
a right-handed basis (the {\it right hand rule} in the terminology of \cite{sasaki14}) is obtained consisting of the gradient of the entropy, $\nabla s$, the rotational component of the velocity, ${\bf v}_R$, and the vorticity, $\boldsymbol\omega$. We can use \eqref{grads} to rewrite the vorticity ${\boldsymbol\omega}$ in the second equation in \eqref{clebsh} as
\begin{equation}\label{vortnew}
    {\boldsymbol\omega}(s,t)
    =  \frac{1}{s}\left(D(s)^{-1}\,({\bf B}(s)
    -\chi(s) {\bf E}(t))\right) \times {\bf v}_R,
\end{equation}
which, in turn, can be used to find an upper bound on the magnitude $|{\boldsymbol\omega}|$ of the vorticity as
\begin{equation}\label{vortest}
    |{\boldsymbol\omega}(s,t)|
    \leq
   \frac{|{\bf v}_R|}{s} ||D(s)^{-1}||(|{\bf B}(s)| + ||\chi(s)|||{\bf E}(t))|)
\end{equation}
where $||\cdot||$ stands for a matrix norm.

The boundary layer also undergoes strong wind shear, which leads to the development of a quasi-two-dimensional turbulent boundary layer. The vorticity in the moisture-rich boundary layer creates potential vorticity \cite{dutton76, holton19} patches. The interaction between the entropic source and the entropic sink leads to a thermodynamic non-equilibrium. This, along with the surface heat flux into the quasi-two-dimensional turbulent boundary layer combined with updrafts below the supercell wall cloud \cite{parker23}, combine to tilt and stretch the vortex lines and unravel their cusps, possibly leading to tornadogenesis.

\section{Considerations of the Gross--Pitaevskii Equation}
\label{sec:Gross--Pitaevskiis}
We conclude the main ideas of this paper by considering the Gross--Pitaevskii equation, written in the dimensionless form \cite{bao13, khesin18},
\begin{equation}
    \label{Gross--Pitaevskiie}
    i \, \frac{\partial {\Psi}}{\partial t} = - \Delta {\Psi}
    + \beta  \vert {\Psi}  ({\bf x}, t) \vert^2 {\Psi}
    + V ({\bf x}) {\Psi},
\end{equation}
where the unknown ${\Psi}({\bf x},t)$ typically represents a wave function, $V({\bf x})$ is an external potential, and $\beta>0$. Notice the similarity to the cubic nonlinear Schr\"odinger equation \eqref{shr2}, particularly the cubic nonlinearity. Note also that the term $I[\psi(\gamma,t)]$ in \eqref{shr2} and defined in \eqref{Iterm} is replaced by a simpler linear term $V({\bf x}){\Psi}$. Equation \eqref{Gross--Pitaevskiie} has several properties that may be useful in studying tornadogenesis. For example, it has been shown that its statistics include the $-5/3$ power law for the inverse energy cascade in two-dimensional turbulent flows \cite{muller24}. Hence \eqref{Gross--Pitaevskiie} can be used to model two-dimensional turbulence and we can try to use it to model quasi-two-dimensional turbulence in the boundary layer. Additionally, a particular modification of the Rankine combined vortex is a solution to \eqref{Gross--Pitaevskiie} \cite{barenghi24}. Finally, we point out that we are not pursuing a quantum mechanical analogy and do not claim any quantum mechanical effects. Instead, we are only focusing on the mechanical properties of \eqref{Gross--Pitaevskiie} and its solutions.

If we rewrite the wave function ${\Psi}$ in terms of its magnitude and phase $\phi$,
\begin{equation*}
    {\Psi}  ({\bf x}, t)
    = \vert {\Psi}  ({\bf x}, t) \vert
    e^{i \phi ({\bf x}, t)},
\end{equation*}
and use the identification
\begin{equation*}
    \rho({\bf x},t)
    =
    \vert {\Psi}({\bf x},t)\vert^2
    \quad\text{ and }\quad
    {\bf v}({\bf x},t)=\nabla\phi({\bf x},t),
\end{equation*}
then the Madelung transform of \eqref{Gross--Pitaevskiie} leads to a continuity equation and a modified Euler equation with $\rho$ and $\bf v$ playing the usual roles of mass density and velocity, respectively \cite{roitberg21,barenghi24}. In one spatial dimension, the Madelung transform agrees with the Hasimoto transform \eqref{hasimoto} \cite{khesin18}. In fact, the Madelung and Hasimoto transforms establish an equivalence of the Gross--Pitaevskii and the nonlinear Schr\"{o}dinger equation classes to Newton's and fluid flow equations including the Euler and the Navier--Stokes equations \cite{khesin18}, and many features of turbulent vortices have been obtained using numerical simulations of the solutions of the Gross--Pitaevskii equation \cite{barenghi24}. The hierarchy of the cubic nonlinear Schr\"{o}dinger equations, including \eqref{Gross--Pitaevskiie}, is studied in \cite{chen14}. As seen above, solutions of \eqref{Gross--Pitaevskiie} can be shown, after an application of the Madelung transform, to satisfy the Euler equations \cite{roitberg21}, and, vice versa, the Euler equations can be turned to \eqref{Gross--Pitaevskiie} using the inverse Madelung transform \cite{strong12}. This equivalence between the Euler equations and the Gross--Pitaevskii equation indicates that (similarly to densely packed sources of rotation perpetrated by the Bose--Einstein statistical distribution) the turbulent interaction includes a ``discrete-like'' component \cite{khesin18,barenghi24}.

Using the usual approach of separation of variables, stationary solutions of \eqref{Gross--Pitaevskiie} of the form
\begin{equation}
    \label{separ}
    {\Psi}  ({\bf x}, t) = Z ({\bf x}) e^{ - i \mu t}
\end{equation}
can be sought. Substituting \eqref{separ} into \eqref{Gross--Pitaevskiie}, we obtain the problem
\begin{equation*}
    \mu \, Z ({\bf x}) = - \Delta Z ({\bf x})
    + \beta  \vert Z  ({\bf x}) \vert^2 Z ({\bf x})
    + V ({\bf x}) Z ({\bf x}),
\end{equation*}
which is to be solved for the eigenvalue $\mu$ and the eigenfunction $Z({\bf x})$, which then fully determine $\Psi({\bf x},t)$ \cite{davis01, davis01b}.

In light of the above connections between the Schr\"odinger equations and the Gross--Pitaevskii equation, it appears plausible that applications of the variants of \eqref{Gross--Pitaevskiie} could possibly lead to broadening the type of vortex patches in our consideration, i.e., the narrow furrows of the pre-turbulent flow, to larger patches in a more fully developed quasi-two-dimensional turbulence.

\section{Summary}
\label{sec:summary}

In this paper we discussed the mathematical, statistical mechanical, and thermodynamical concepts that allow us to better understand the development of a three-dimensional vortex from an originally mostly non-turbulent flow. Surface friction and vertical wind shear lead to the development of horizontal vortex filaments, which can later develop into more vertical coherent structures such as cusps and furrows, exhibiting apparent hairpin legs \cite{markowski24,adrian07,bernard11,bernard19}.

Since these coherent structures occur in a shallow region near the surface (see Fig.~\ref{fig:bernard2}), their behavior can be modeled and understood using quasi-two-dimensional models. In two-dimensional and therefore also quasi-two-dimensional models, energy cascades to larger scales, further supporting the development of structures or patches discussed in \cite{fischer24}. Such structures are then pulled into the updraft of the supercell flow~\cite{parker23}, get further tilted, unraveled, and stretched in the vertical direction \cite{markowski24} (see Fig.~\ref{fig:coherent}), and a three-dimensional model is required. When the cyclostrophic balance between the pressure gradient force and the centrifugal force is disrupted due to the surface friction, these stretched filaments collapse into narrow, very intense filaments. Such filaments, further stretched by the updraft, then transfer energy to the surrounding flow, possibly leading to or supporting tornadogenesis \cite{belikdokken23,belikdokkenpotvinscholzshvartsman17,belikdahldokkenpotvinscholzshvartsman18}.

A mathematical explanation of the development of cusp-like features from horizontal vortex filaments is provided via nonlinear Schr\"odinger equations such as \eqref{shr1} and \eqref{shr2}. In particular, the nonlocal interaction term $I[\psi]$ in \eqref{shr2} is shown in \cite{kleinmajda912} to be responsible for the development of cusps in an initially straight, helically perturbed, vortex filament. These models are rigorously derived in \cite{kleinmajda91,kleinmajda912} under appropriate and reasonable mathematical assumptions. The models' assumptions cease to be satisfied as the cusps are formed; however, when such cusps are formed, their behavior can be approximated by using models appropriate for collections of vortices, such as a two-dimensional vortex gas model or a quasi-two-dimensional model for parallel vertical vortices.

The results of the numerical simulations in \cite{markowski24} appear to be consistent with those in \cite{bernard11} in that $\lambda_2$-isosurfaces in the boundary layer of the environmental air mass align with the direction of the flow, and transverse vortex filaments develop cusps and potentially more complicated shapes between these ``hairpin legs.'' It is important to point out that the horizontal hairpin legs do not get tilted into the vertical. Instead, the transverse vortex filaments with cusps appear to be tilted by a combination of potential vorticity in the boundary layer and the pressure drop in the mesocyclone. These filaments then get pulled into the updraft and appear to significantly contribute to tornadogenesis.

We have addressed the notion of entropic balance studied by Sasaki \cite{sasaki14}, in which regions of entropic sources and sinks are identified and studied in the context of the entire supercell storm. The interaction of these sources and sinks leads to a thermodynamic non-equilibrium in the supercell. In the context of tornadogenesis, this theory can be applied to the boundary layer. The entropic sink is the evaporatively cooled boundary layer in the forward flank or rear flank of the supercell. The entropic source is the boundary layer, laden with moisture from northward advection of water vapor from the Gulf of Mexico and the absorption of water vapor from crops undergoing evapotranspiration. The close proximity of the entropic source and the entropic sink determines the size of the entropy gradient, and their interaction, together with other factors such as strong wind shear or surface heat flux, leads to the generation of (potential) vorticity in this quasi-two-dimensional layer that appears to be essential for tornadogenesis. Combined with updrafts below the supercell wall cloud, this mechanism leads to tilting, unraveling, and stretching of the cusps along vortex lines in this region, contributing to tornadogenesis.

Finally, we list some questions to consider for future investigation. What are the thresholds for surface winds to produce quasi-two-dimensional turbulence with robust vortex patches? Prior to tornadogenesis, how does the near-storm thermodynamic environment affect how close the vortex patches get to the triple point, i.e., the point where the surface air from the forward-flank and the rear-flank downdrafts join with the surface environmental air? What roles do the ground flux and the moisture flux in the non-equilibrium thermodynamics play in tornadogenesis? Is there a way to continuously perturb the linear local term into a nonlocal term in the cubic nonlinear Schr\"{o}dinger equation in such a way that the evolution of the solutions reflects the increasing turbulence? If there is a homotopy of evolution from the Schr\"{o}dinger to the Gross--Pitaevskii equation mentioned above, can we then find entropy purely based on a time-dependent wave function without using a quantum analogy?









\bibliography{thermod}
\bibliographystyle{abbrv}

\end{document}